\documentclass[
reprint, 
superscriptaddress,
amsmath, 
amssymb, 
aps, 
prb, 
longbibliography,
]{revtex4-2}

\usepackage[dvipsnames]{xcolor}
\definecolor{mblue}{RGB}{31, 119, 180}
\usepackage{hyperref}
\hypersetup{backref,pdfpagemode=FullScreen,colorlinks=true,breaklinks,urlcolor=mblue,linkcolor=mblue,citecolor=mblue}
\usepackage{physics, braket, bm, amsthm, amsmath, amssymb, mathrsfs, graphicx, dcolumn}
\usepackage[linesnumbered,ruled,lined]{algorithm2e}
\usepackage[]{ulem}

\newcommand{\tabincell}[2]{\begin{tabular}{@{}#1@{}}#2\end{tabular}}

\begin{document}

\preprint{12234}

\title{Neural network quantum state with proximal optimization: a ground-state searching scheme based on variational Monte Carlo}


\author{Feng Chen}
\email{chenfengwhu@gmail.com}
\affiliation{State Key Laboratory of Low Dimensional Quantum Physics, Department of Physics, Tsinghua University, Beijing 100084, China}

\author{Ming Xue}
\email{mxue@nuaa.edu.cn}
\affiliation{College of Physics, Nanjing University of Aeronautics and Astronautics, Nanjing 211106, China}
\affiliation{Key Laboratory of Aerospace Information Materials and Physics (NUAA), MIIT, Nanjing 211106, China}
\affiliation{State Key Laboratory of Low Dimensional Quantum Physics, Department of Physics, Tsinghua University, Beijing 100084, China}


\date{\today}
\begin{abstract}
Neural network quantum states (NQS),
incorporating with variational Monte Carlo (VMC) method,
are shown to be a promising way to investigate quantum many-body physics.
Whereas vanilla VMC methods perform one gradient update per sample, 
we introduce a novel objective function with proximal optimization (PO) 
that enables multiple updates via reusing the mismatched samples.
Our VMC-PO method keeps the advantage of the previous importance sampling gradient optimization algorithm
[\href{https://doi.org/10.1103/PhysRevResearch.2.012039}{L. Yang, {\it et al}, Phys.\;Rev.\;Research\;{\bf 2}, 012039(R)(2020)}]
that efficiently uses sampled states.
PO mitigates the numerical instabilities during network updates, which is similar to 
stochastic reconfiguration (SR) methods, but achieves an alternative and simpler implement with lower computational complexity.
We investigate the performance of our VMC-PO algorithm for ground-state searching
with a 1-dimensional transverse-field Ising model and 2-dimensional Heisenberg antiferromagnet on a square lattice,
and demonstrate that the reached ground-state energies are comparable to state-of-the-art results.
\end{abstract}
 

\maketitle

\section{Introduction}
Finding efficient representation for quantum states is a long-sought pursuit in the realm of quantum many-body physics, 
which is of great importance in a wide range of condensed-matter physics, nuclear physics, and quantum chemistry problems.
Variational {\it Ans\"atzes} inspired from physical insights, 
such as matrix product states concerning the fact that most physically interesting states 
in one dimension are area-law entangled, are extensively and successfully explored in 
representing quantum many-body states~\cite{21rmpCirac}.
Sophisticated computational methods like quantum Monte Carlo techniques~\cite{sandvik2010computational},
tensor network ansatzes~\cite{SCHOLLWOCK201196, orus2014practical} and the dynamical mean-field theory~\cite{georges96rmp},
present commendable efforts in addressing strongly correlated many-body systems,
but each of them can only address specific classes of problems efficiently, and many remain yet to be figured out.

Artificial neural networks (ANN) as a class of models for parametrization are known 
to be versatile, powerful, and scalable, making them ideal to tackle large and highly complex tasks~\cite{19rmpml}.
Their strong power of expressivity and highly efficient optimization aided by 
scalable learning algorithms and GPU acceleration,
are invaluable resources for solving quantum many-body physics~\cite{carleo2017solving,torlai2018neural,huang2022provably,dawid2022}.
Combining insights from machine learning and computational quantum physics,
ANNs have been successfully adapted to circumvent the curse of dimensionality on simulating quantum many-body systems on classical computers. 

The variational {\it Ans\"atze} of quantum wave functions parametrized by ANNs,
known as {\it neural-network quantum states} (NQS)~\cite{carleo2017solving},
shows close connection with {\it tensor network states} (TNS) that widely used in quantum many-body studies,
and is capable of representing quantum many-body states with fewer parameters compared to its TNS counterpart~\cite{Chen18PRB}.

Since the NQS {\it Ans\"atze} was first proposed by Carleo and Troyer~\cite{carleo2017solving}, a variety of ANN architectures
are designed and investigated to solve the quantum many-body problems, such as
restricted Boltzman machine~\cite{carleo2017solving,carleo2018constructing,Pastori19prb,nomura2021}, 
convolutional neural networks~\cite{Choo19prb,Markus20prl,bukov2021}, 
deep autoregressive models~\cite{Sharir20deep,wu21prr}, and variational autoencoders~\cite{rocchetto2018learning}, to just name a few. 

The optimization of the NQS {\it Ans\"atzes} are usually done by neural-network-based variational Monte Carlo (VMC)
with stochastic samples according to Born probability distributions. 
However, a large amount of samples is demanded for accurate evalvation of the optimization objective.
Inefficient sampling hinders the process of optimization, thus shading the expressivity of a variational NQS {\it Ans\"atze}.
One of the usual ways to efficiently generate these samples is Markov-chain Monte Carlo (MCMC)~\cite{press2007numerical}, an importance sampling method.
However, these samples are discarded after a single use in the conventional neural-network-based VMC methods,
even though the {\it Ans\"atze} is not substantially updated.
Poineering study of Ref.~\cite{yang2020} have adopted an important sampling gradient optimization (ISGO) approach
to modify the conventional VMC objective function, 
which introduces additional inner loop to reuse the mismatched samples before requiring new samples from MCMC.
The efficacious reuse of samples hence improve the computational efficiency of training deep networks at the expense of robustness. 
A natural-gradient-descent based approach, stochastic reconfiguration (SR) or imaginary time evolution~\cite{sorella1998,Sorella01prb}, by contrast, is relatively stable but computational sophisticated  and treats the samples in a disposable way.

This work seeks to improve the VMC-ISGO approach by introducing a proximal optimization (PO) algorithm,
denoted as VMC-PO, that simultaneously attains the sample efficiency and reliable performance of SR,
while using only first-order optimization. 
We propose an  objective with PO containing clipped probability ratios that
forms a pessimistic estimate and phase constraints as penalties.
The constraints on variant of the wave function will improve the stability 
of optimizing parameters within the inner loop.

The rest of this paper is organized as follows.
Section \ref{sec_NN} is devoted to introduce the neural network architectures of our variational NQS {\it Ans\"atze}.
In Sec.~\ref{sec_opt}, we revisit the conventional VMC and VMC-ISGO approaches and derive our optimizaition algorithm (VMC-PO) step by step. 
Then in Sec.\,\ref{sec_exp}, we benchmark our optimization approaches on solving the 
ground states of a spin-$\frac{1}{2}$ transverse field Ising model (TFIM) in 1D lattice
and a 2D square lattice with spin-$\frac{1}{2}$ Heisenberg antiferromagnet (HAFM).
Finally in Sec.\,\ref{sec_con}, we conclude with a discussion and outlook.
\begin{figure} 
    \includegraphics[width=\columnwidth]{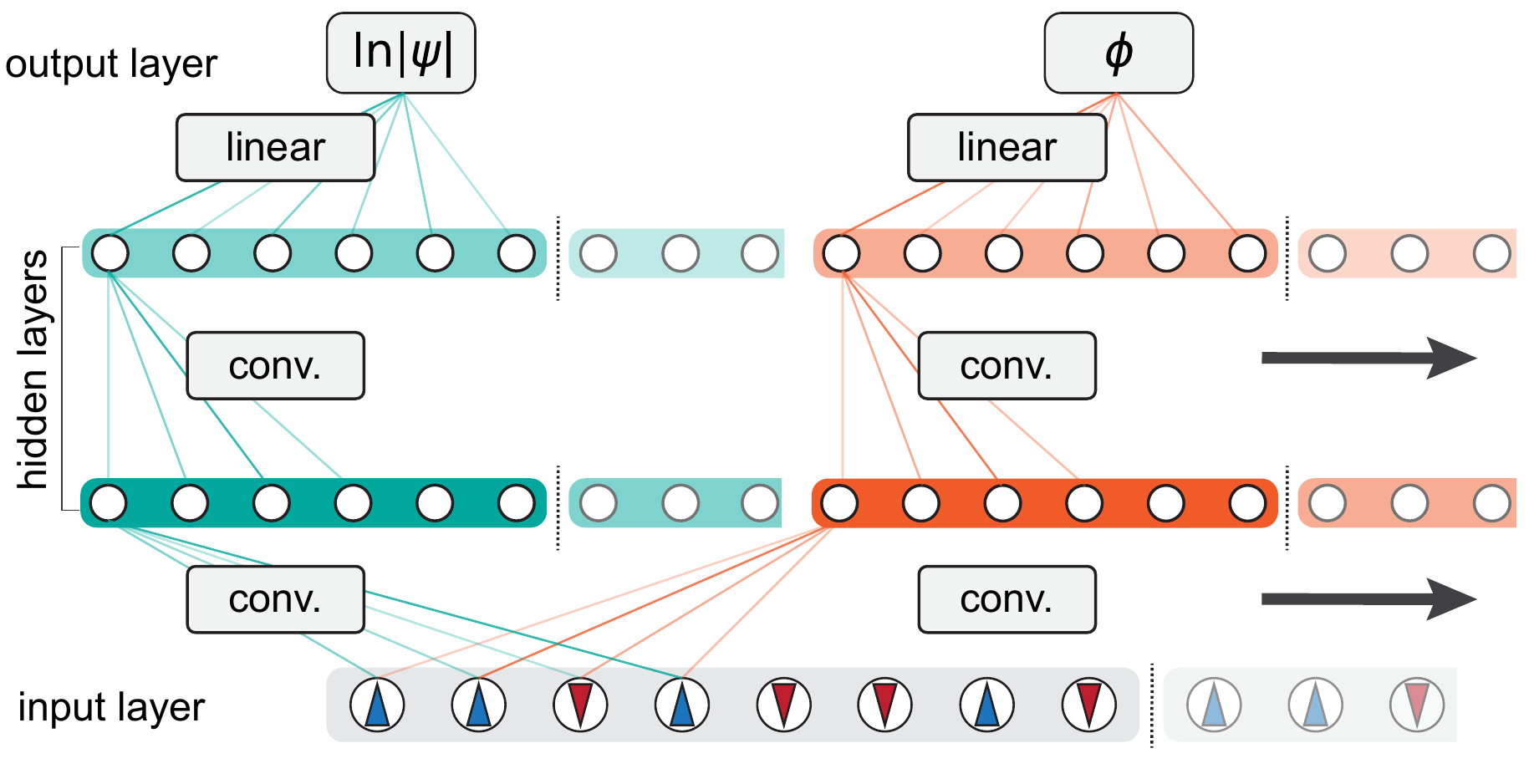}
    \caption{Schematic CNNs with two hidden layers and single feature in a 1D system containing
    spin-$\frac{1}{2}$ particles ($D_{\rm p} = 2$).
    The size of convolutional kernel is $K=4$.
    Periodic boundary condition (pbc) is considered via periodic padding on the input and hidden layers
    (right hand side of dotted vertical lines).
    The green (orange) lines show layer-to-layer couplings in the amplitude (phase) network,
    which are repeated for all neurons (white circles).
    The black arrows show the direction of the translational operations of the convolutional kernels.
    }
    \label{fig_cnn}
\end{figure}

\section{Network architectures} \label{sec_NN}
The NQS {\it Ans\"atze} is parametrized with real variational parameters $\bm \theta$: 
$\langle {\bm s}|\Psi_{\bm \theta}\rangle\equiv\Psi_{\bm \theta}(\bm s)=|\Psi_{\bm \theta}(\bm s)|e^{i\Phi_{\bm \theta}(\bm s)},$
where the network input $\bm s$ denotes the physical basis for sampling.
It is convenient to work directly with the logrithmic wavefunction~\cite{carleo2017solving, Bukov21learning}. 
Therefore the complex-valued representation can be seperated into two decoupled networks with real-valued parameters:
\begin{equation}
    \ln\Psi(\bm s; {\bm \theta}) = \mathcal{A}_{\bm\theta}({\bm s}) + i\Phi_{\bm\theta}({\bm s}),
\end{equation}
i.e., the amplitude network $\mathcal{A}_{\bm \theta}({\bm s}) \equiv \ln |\Psi_{\bm \theta}(\bm s)|$ 
and the phase network $\Phi_{\bm\theta}({\bm s})$.
The variational parameters $\bm\theta$ are determined by minimizing the energy functional, 
$E = \braket{\Psi|H|\Psi}/\braket{\Psi|\Psi}$, within the basis state $\{|\bm s^{(m)}\rangle\}$ 
(where $m$ the sample index).
Each sampled state $|\bm s^{(m)}\rangle$ is encoded as an input tensor ${s}^{(m)}_{{\bf r}, f}$,
where the indexes ${\bf r}$ and $f$ denote the lattice site and the local spin state.

In a 1-dimensional (1D) lattice spin system, two identical $L$-layer 1D CNNs are
implemented to respectively represent the amplitude and phase parts (as illustrated in Fig.~\ref{fig_cnn}).
Take the amplitude network $\mathcal{A}_\theta$ as an instance:
An input state tensor $s^{(m)}_{i, f}$($i=1,\ldots, N$) is fed into the first hidden layer 
with one-hot coding for local spin states~\cite{yang2020},
and the $D_{\rm p}$-dimensional spin degree of freedom is regarded as the channels of input features.
The feature map $f$ of the first hidden layer is then calculated by translating the
convolutional kernel along the spatial dimension, 
\begin{equation}
    \label{eq:cnn1d_1}
    A^{[1]}_{i,f} = \sigma\left(\sum^K_{k=1}\sum^{D_{\rm p}}_{f^\prime=1} W^{[1]}_{k,f',f}s^{(m)}_{i+k,f'} + b^{[1]}_{f}\right),
\end{equation}
where $K$ is the kernel size of the filter.
${\bm W}$ and $\bm b$ are weights of the kernel and bias, respectively,
which are represented by trainable network parameters.
$\sigma(\cdot) = {\rm selu}(\cdot)$ is the nonlinear activation function for enhanced expressivity.
Boundary conditions are considered as appropriate padding operations on the feature maps, 
e.g., periodic padding operation for the periodic boundary condition (pbc)
and zero padding operation for the open boundary condition (obc).

Deeper structures are built by feeding a former hidden layer,
which are convoluted by kernels, into a next level layer.
Specifically, a neuron located in lattice-site $i$ of the feature map $f$ 
in a given convolutional layer $l+1$ is connected to the outputs of neurons in previous layer $l$ by,
\begin{align}
    \label{eq:cnn1ds}
    A^{[l+1]}_{i,f} = \sigma\left(\sum^K_{k=1}\sum^{F}_{f'=1} W^{[l+1]}_{k,f',f}A^{[l]}_{i+k,f'} + b^{[l+1]}_{f}\right),
\end{align}
wherein $l=1,2,...,L-1$.
The amplitude of state $|\bm s^{(m)}\rangle$ is obtained after summing the spatial
and feature dimensions of the last hidden layer, 
\begin{equation}
    \mathcal{A}_{\bm \theta}\left({\bm s}^{(m)}\right) = \sum^F_{f=1} a_f\sum_{i=1}^N A^{[L]}_{i,f},
    \label{eq:logpihi_1d}
\end{equation}
where $\{a_f\}$ the weights of the feature map $f$ and are trainable as well, 
hence $\bm \theta\equiv\{\bm W, \bm a, \bm b\}$ represents all the trainable parameters.


To consider the lattice symmetries in a 2D lattice system, we use group convolutional neural netwoks (GCNN)~\cite{cohen2016,roth2021,netket2022}, which generalizes the CNN by acting over a discrete group $G$.
The wave function represented by a GCNN is symmetric over the group $G$, leading to improved accuracy of ground state energy estimation due to the physical prioirs encoded.
For an input state $s^{(m)}_{{\bf r},f}$ in a 2D spatial system with $N\times N$ lattice sites, the feature maps of the hidden layers in the amplitude network are given by,
\begin{eqnarray}
    \label{eq:cnn2d_1}
    A^{[1]}_{{\bf r},g,f} &=& \sigma\left(\sum_{\bf k}\sum^{D_{\rm p}}_{f'=1} W^{[1]}_{{\bf k},g^{-1},f',f}s^{(m)}_{{\bf r+k},f'} + b^{[1]}_{f}\right),\nonumber\\
    A^{[l+1]}_{{\bf r},g,f} &=& \sigma\left(\sum_{\bf k}\sum_{h\in G'}\sum^{F}_{f'=1} W^{[l+1]}_{{\bf k},{h^{-1}g},f',f}A^{[l]}_{{\bf r+k},h,f'} + b^{[l+1]}_{f}\right),\nonumber
\end{eqnarray}
where $l=1,2,3,...,L-1$. 
The feature map is hence indexed by group element $g$ besides spatial position $\bf r$ and feature $f$, where $g\in G'$ and $G'$ is a subgroup of $G$ excluding translation. 


For numerical stability, 
the summing rules we defined in the GCNN~\cite{Bukov21learning} for amplitude and phase networks are different from Eq.~\eqref{eq:logpihi_1d},
\begin{align}
    \ln |\Psi| = \ln&\big[\sum_{{\bf r},g,f} \exp(A^{[L]}_{{\bf r}, g, f}) \big], \nonumber \\
          \Phi = \arg&\big[ \sum_{{\bf r},g,f} \exp(i\Phi^{[L]}_{{\bf r}, g, f}) \big].
    \label{eq:logpihi_2d}
\end{align}

\section{optimization approaches} \label{sec_opt}
\begin{figure}[!b]
    \includegraphics[width=\columnwidth]{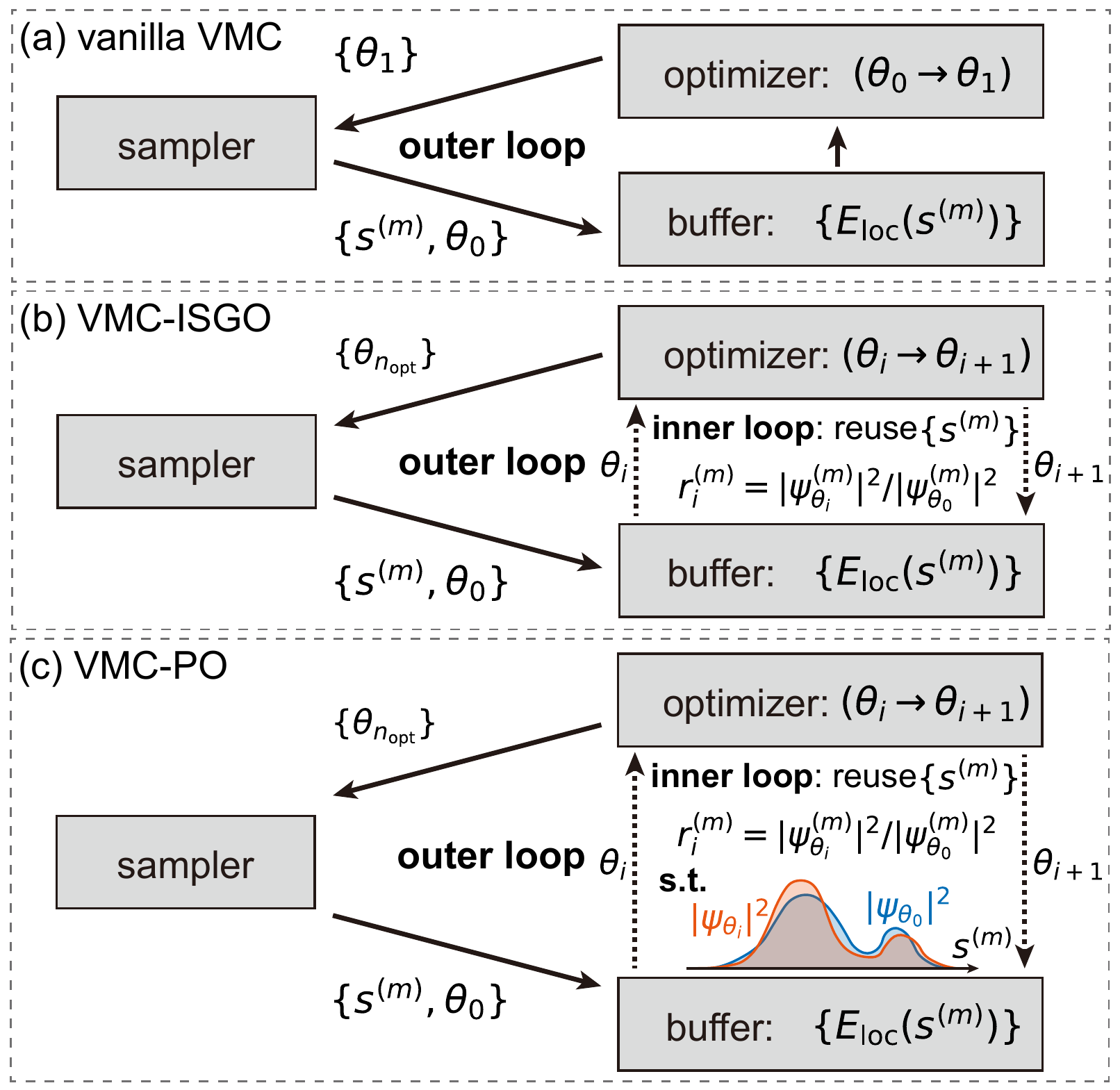}
    \caption{The flowchart comparing (a) the vanilla VMC, (b) the VMC-ISGO and (c) our VMC-PO algorithm.
    The sampler generates a set of $M$ samples $\{\bm s^{(m)}\}$ based on a Markov chain, 
    $\bm s^{(1)}\rightarrow \bm s^{(2)}\rightarrow...\rightarrow\bm s^{(M)}$. 
    Local energies $E_{\rm loc}$ based on the sampled states and parameters $\bm\theta$ are computed in the buffer.
    Network parameters $\bm\theta$ are then updated in the optimizer according to the gradients 
    of the defined objective function $\mathcal{L}$ with respect to $\bm\theta$. 
    The cycle index of outer loop (black solid lines) is {\it epoch} and the one of inner loop (black dotted lines) is $n_{\rm opt}$.
    }\label{fig_algos}
\end{figure}

The original objective function of VMC approach is the energy functional.
It is estimated from a set of $M$ samples $\{\bm s^{(m)}\}$ that are sampled
according to the Born probability density $P(\bm s)\propto|\Psi_{\bm\theta}(\bm s)|^2$.
Each sample $\bm s^{(m)}$ is used to calculate a local energy
$E_{\rm loc}(\bm s^{(m)}) = \langle \bm s^{(m)}|H|\Psi\rangle/\langle \bm s^{(m)}|\Psi\rangle = E^R_{\bm s^{(m)}} + iE^I_{\bm s^{(m)}}$. 
The objective function is the statistical average of all $E_{\rm loc}$ calculated using the samples $\{\bm s^{(m)}\}$,
\begin{equation}
    \label{eq:vmc}
    \mathcal{L}^{\rm VMC}(\bm\theta) = \frac{1}{M} \Re\big\{ \sum_m E_{\rm loc}(\bm s^{(m)}) \big\}.
\end{equation}
Network parameters $\bm\theta$ are updated according to gradient descent, 
$\bm\theta' \leftarrow \bm\theta + \alpha\cdot\partial_{\bm\theta} \mathcal{L}(\bm\theta)$,
where $\alpha$ is learning rate of optimizer.
The optimizer we choose here is {Adam}~\cite{kingma2014adam}, 
an efficient variant of the stochastic gradient descent algorithm.
The derivative of objective function \eqref{eq:vmc} with respect to variational parameters is given by,
\begin{eqnarray}
    \label{eq:diff_vmc}
    \partial_{\bm\theta} \mathcal{L}^{\rm VMC}(\bm\theta) &=& \frac{1}{M}\Re\big\{
        \sum_{m}\!\! E_{\bm s^{(m)}}^R\partial_{\bm\theta}\mathcal{A}(\bm s^{(m)})\nonumber\\
        &&+\sum_m E^I_{\bm s^{(m)}}\!\partial_{\bm\theta}\Phi(\bm s^{(m)})\nonumber\\
        &&- \frac{1}{M}\sum_{m}\! E_{\bm s^{(m)}}^R\sum_{m}\partial_{\bm\theta}\mathcal{A}(\bm s^{(m)})\big\}.
\end{eqnarray}
After the parameters $\bm\theta$ are updated,
the wave function changes from $\Psi_{\bm\theta}$ to $\Psi_{\bm\theta'}$. 
As illustrated in Fig.~\ref{fig_algos}(a), 
a new set of samples based on the updated wave function $\Psi_{\bm\theta'}$ 
is demanded to estimate the objective function \eqref{eq:vmc} for the subsequent iterations.
The iterations here are called outer loop [Fig. 2(a)].
The previously sampled states $\{|\bm s\rangle\}$ based on $\Psi_{\theta}(\bm s)$ will be discarded.

To utilize the mismatched samples, a previous study~\cite{yang2020} develop an efficient ISGO method.
It renormalizes the distribution of those mismatched samples to $|\Psi_{\bm\theta'}|^2$ by multiplying local energies with importance sampling factors.
The surrogate objective function to minimize is modified in ISGO as,
\begin{equation}
    \label{eq:isgo}
    \mathcal{L}^{\rm ISGO}(\bm\theta) = \frac{\mathcal C}{M} \Re\big\{\sum_m r^{(m)} E_{\rm loc}(\bm s^{(m)})\big\},
\end{equation}
where $E_{\rm loc}$ are the local energies under the last updated network parameters $\bm\theta'$
based on the old samples from the last epoch,
and $r^{(m)} = |\Psi_{\bm\theta'}(\bm s^{(m)})|^2/|\Psi_{\bm\theta}(\bm s^{(m)})|^2$ 
denotes the probability ratio from importance sampling,
the normalization factor $\mathcal C$ is approximately calculated by assuming ${\mathcal C}\sum_m r^{(m)}/M=1$.
The gradient of Eq.~\eqref{eq:isgo} is thus given by,
\begin{widetext}
\begin{equation}
    \label{eq:diff}
    \partial_{\bm\theta} \mathcal{L}^{\rm ISGO}(\theta) = \frac{\mathcal C}{M}\Re\left\{
        \sum_{m} r^{(m)}\!\!\left[E_{s^{(m)}}^R\partial_{\bm\theta}\mathcal{A}(\bm s^{(m)})+ E^I_{s^{(m)}}\partial_\theta\Phi(\bm s^{(m)})\right]
        - \frac{\mathcal C}{M}\sum_{m}r^{(m)}E_{s^{(m)}}^R\sum_{m}r^{(m)}\partial_\theta\mathcal{A}(\bm s^{(m)})\right\},
\end{equation}
\end{widetext}
ISGO introduces an inner loop [see Fig.~\ref{fig_algos}(b)] to update the network 
parameters $\bm\theta$ before requiring new samples,
which leads to efficient use of the $M$ sampled states.

Empirically, this method often leads to destructively large energy updates or overusing of sampled states in a single epoch. 
Reducing cycles of the inner loop and choosing lower learning rate are just palliative treatments.
An alternative approach to avoid the large update and overfitting is implementing constraints on the objective function.
One natural choice of constraint is based on the Fubini-Study metric,
\begin{equation}
    \label{eq:dfs}
    {\mathcal D}(\Psi_{\bm\theta'}, \Psi_{\bm\theta}) = \arccos\sqrt{\frac{\langle \Psi_{\bm\theta}|\Psi_{\bm\theta'}\rangle\langle 
        \Psi_{\bm\theta'}|\Psi_{\bm\theta}\rangle}{\langle \Psi_{\bm\theta}|\Psi_{\bm\theta}\rangle\langle \Psi_{\bm\theta'}|\Psi_{\bm\theta'}\rangle}},
\end{equation}
which describes the distance between two quantum states $\Psi_{\bm\theta'}, \Psi_{\bm\theta}$. 
The optimal parameters $\bm\theta^*$ are hence determined by minimizing the objective function subject to a constraint on the size of update,
\begin{align}
    \label{eq:trpo}
    \bm\theta^* = \mathop{\rm argmin}_{\bm\theta}~\mathcal{L}^{\rm ISGO}(\bm\theta),\;
    \text{subject to}\;\,{\mathcal D}^2(\Psi_{\bm\theta'}, \Psi_{\bm\theta}) \leqslant \delta. \nonumber
\end{align}
If the variation in the parameters is small, say $\bm\theta'_i=\bm\theta_i +{\rm d}\bm\theta_i$,
the updated wave function can be written as 
$\Psi_{\bm\theta'} = \Psi_{\bm\theta} + \partial_{\bm\theta}\Psi_{\bm\theta}\cdot {\rm d}\bm\theta 
= \Psi_{\bm\theta} + \Psi_{\bm\theta}\bm{\mathcal{O}} \cdot{\rm d}{\bm\theta}$,
where $\bm{\mathcal{O}} \equiv \partial_{\bm\theta}\ln\Psi_{\bm\theta} = \partial_{\bm\theta}\ln|\Psi_{\bm\theta}| + i\partial_{\bm\theta}\Phi_{\bm\theta}$ are the logarithmic derivates.
The square of the Fubini-Study metric then can be approximately written as,
\begin{equation}
    \label{eq:dfs2}
    {\mathcal D}^2(\Psi_{\bm\theta'}, \Psi_{\bm\theta}) \approx \sum_{ij} {\rm d}\theta_i {\rm d}\theta_j \left[ \langle{\mathcal O}^\dagger_i{\mathcal O}_j\rangle - \langle{\mathcal O}^\dagger_i\rangle \langle{\mathcal O}_j\rangle \right].
\end{equation}
We also expand the surrogate objective function Eq.\,\eqref{eq:isgo} near $\bm\theta$,
and the target becomes finding optimal ${\rm d}\bm\theta^*$ to minimize the erengy,
\begin{align}
    \label{eq:trpo2}
    &{\rm d}\bm\theta^* = \mathop{\rm argmin}_{{\rm d} \bm\theta}~[\mathcal{L}^{\rm ISGO}(\bm\theta)+\partial_{\bm\theta} \mathcal{L}^{\rm ISGO}(\bm\theta){\rm d}\bm\theta], \nonumber \\
    &{\rm subject~to}~~\sum_{ij} {\rm d}\theta_i {\rm d}\theta_j \left[ \langle{\mathcal O}^\dagger_i{\mathcal O}_j\rangle - \langle{\mathcal O}^\dagger_i\rangle \langle{\mathcal O}_j\rangle \right] \leqslant \delta,
\end{align}
where $\langle \cdot\rangle$ means statistical average upon the samples.
According to natural gradient descent~\cite{amari1998}, the solution of this equation is given by,
\begin{equation}
    \label{eq:trpo3}
    \partial_{\bm\theta} \mathcal{L}^{\rm ISGO}(\bm\theta) + \frac{1}{\alpha} \sum_j S_{ij} {\rm d}\theta_j = 0,
\end{equation}
or
\begin{equation}
    \label{eq:trpo4}
    {\rm d}\bm\theta = -\alpha{\bm S}^{-1}\cdot\partial_{\bm\theta} \mathcal{L}^{\rm ISGO},
\end{equation}
where $S_{ij} = \langle{\mathcal O}^\dagger_i{\mathcal O}_j\rangle - \langle{\mathcal O}^\dagger_i\rangle \langle{\mathcal O}_j\rangle$,
called the SR matrix or the Fisher information matrix~\cite{Sorella01prb}.
The parameters $\bm\theta$ are updated according to $\bm\theta \leftarrow \bm\theta + {\rm d}\bm\theta$.
When $r^{(m)} = 1$, the first cycle in the inner loop, $\mathcal{L}^{\rm ISGO} = \mathcal{L}^{\rm VMC}$,
this method is actually the SR approach or imaginary time evolution~\cite{becca2017quantum}.
However, the calculation of the inverse of the SR matrix, ${\bm S}^{-1}$,
is computational consuming, especially when using a large number of sampled states or complicated networks with lots of parameters.
We hence prefer an alternative way to build the constraints on the objective function Eq.~\eqref{eq:isgo},
which should be numerically more friendly as well.

Inspired by the proximal policy optimization algorithm~\cite{ppo2017} in reinforcement learning,
we purpose a surrogate objective function,
\begin{widetext}
\begin{equation}
    \label{eq:ppo}
    \mathcal{L}^{\rm clip}(\bm\theta) =
    \frac{1}{M} \Re\left\{ \sum_m \max\left[ {\mathcal C} r^{(m)} E_{\rm loc}(\bm s^{(m)}),~ {\mathcal C'} {\rm clip}(r^{(m)}, 1-\epsilon, 1+\epsilon) E_{\rm loc}(\bm s^{(m)}) \right]\right\} -  \frac{\beta}{M}\sum_m \cos(\Delta \phi).
\end{equation}
\end{widetext}
The added constraints on the amplitudes by demanding the ratio $r^{(m)}$ close to 1 [see Fig.~\ref{fig_algos}(c)].
The first term inside the max is the surrogate local energy of ISGO method. 
The second term, ${\rm clip}(r^{(m)}, 1-\epsilon, 1+\epsilon) E_{\rm loc}(\bm s^{(m)})$,
modifies the ISGO local energies by clipping the probability ratio, which removes the incentive for moving $r^{(m)}$ outside of the interval $[1-\epsilon,1+\epsilon]$. 
The normalized factor $\mathcal C'$ of the second term is approximately determined 
according to ${\mathcal C'}\sum_m {\rm clip}(r^{(m)}, 1-\epsilon, 1+\epsilon)/M=1$,
$\epsilon = 0.1$ is the clip ratio.
We then take the maximum of the clipped and unclipped objective,
so the final objective gives an uppper bound on the unclipped objective. 
When the upper bound is decreased, the NQS will become closer to the true ground state.
An effective constraint on the variant of phase is introduced by the last term of Eq.~\eqref{eq:ppo},
$\sum_m \cos(\Delta \Phi) \equiv \sum_m \cos\left[ \Phi_{\bm\theta'}(\bm s^{(m)}) - \Phi_{\bm\theta}(\bm s^{(m)})\right]$,
as a penalty on the objective.
An adaptive penalty coefficient $\beta\in[\beta_{\min}, \beta_{\max}]$ is applied,
and $\beta$ is updated according to the value of $d = \sum_m \cos(\Delta\Phi)/M$, 
\begin{align*}
\beta &\leftarrow\beta/1.5,& &\text{if}\quad d < \beta_{\rm tg_{\min}},\\
\beta &\leftarrow 1.5\cdot\beta,& &\text{if}\quad d > \beta_{\rm tg_{\max}}.
\end{align*}
The updated coefficient $\beta$ is used for the next parameter update (next cycle of the inner loop),
and its inital value is 1. The hyperparameters, 1.5, $\beta_{\rm tg_{\min}}$,
and $\beta_{\rm tg_{\max}}$ are heuritical choices,
but our VMC-PO algorithm is not very sensitive to them 
(details of hyperparameters can be found in Appendix~\ref{app_A}).

The computational complexity of the objective function Eq.\,\eqref{eq:ppo} is on the same order as ISGO method,
which only reqiures the first-order derivatives with respect to the parameters $\bm\theta$.
These derivatives hence can be efficiently evaluated via only single backward propagation 
of the neural netwroks $\mathcal{A}_{\bm\theta}$ and $\Phi_{\bm\theta}$.
In the actual training process, for numerical stability,
we separately train the amplitude and phsae networks.
Specifically, in each training epoch, the phase network $\Phi_{\bm \theta}$ is updated 
before the amplitude network $\mathcal{A}_{\bm\theta}$,
since updating phase network does not change the Born probability distribution.


\section{numerical results} \label{sec_exp}


\subsection{1D Transverse Field Ising Model}
We consider a 1D TFIM with ferromagnetic interaction and periodic boundary condition,
\begin{equation}
\label{ham_tfim}
H = -\sum_i S^i_z S^{i+1}_z + h\sum_i S^i_x,
\end{equation}
where $h$ is the strength of transverse field.
An uniformly initialized NQS is a product state, $\left(\frac{|\uparrow\rangle+|\downarrow\rangle}{\sqrt{2}}\right)^{\otimes N}$, 
which is the ground state of the Hamiltonian \eqref{ham_tfim} when $h\rightarrow-\infty$ 
or the highest excited state when $h\rightarrow\infty$.
We benchmark our algorithm at the quantum critical points $h=\pm0.5$,
which separate the anti-ferromagnetic phase ($|h|<0.5$) and the paramagnetic phase ($|h|>0.5$).

\begin{figure} 
    \includegraphics[width=\columnwidth]{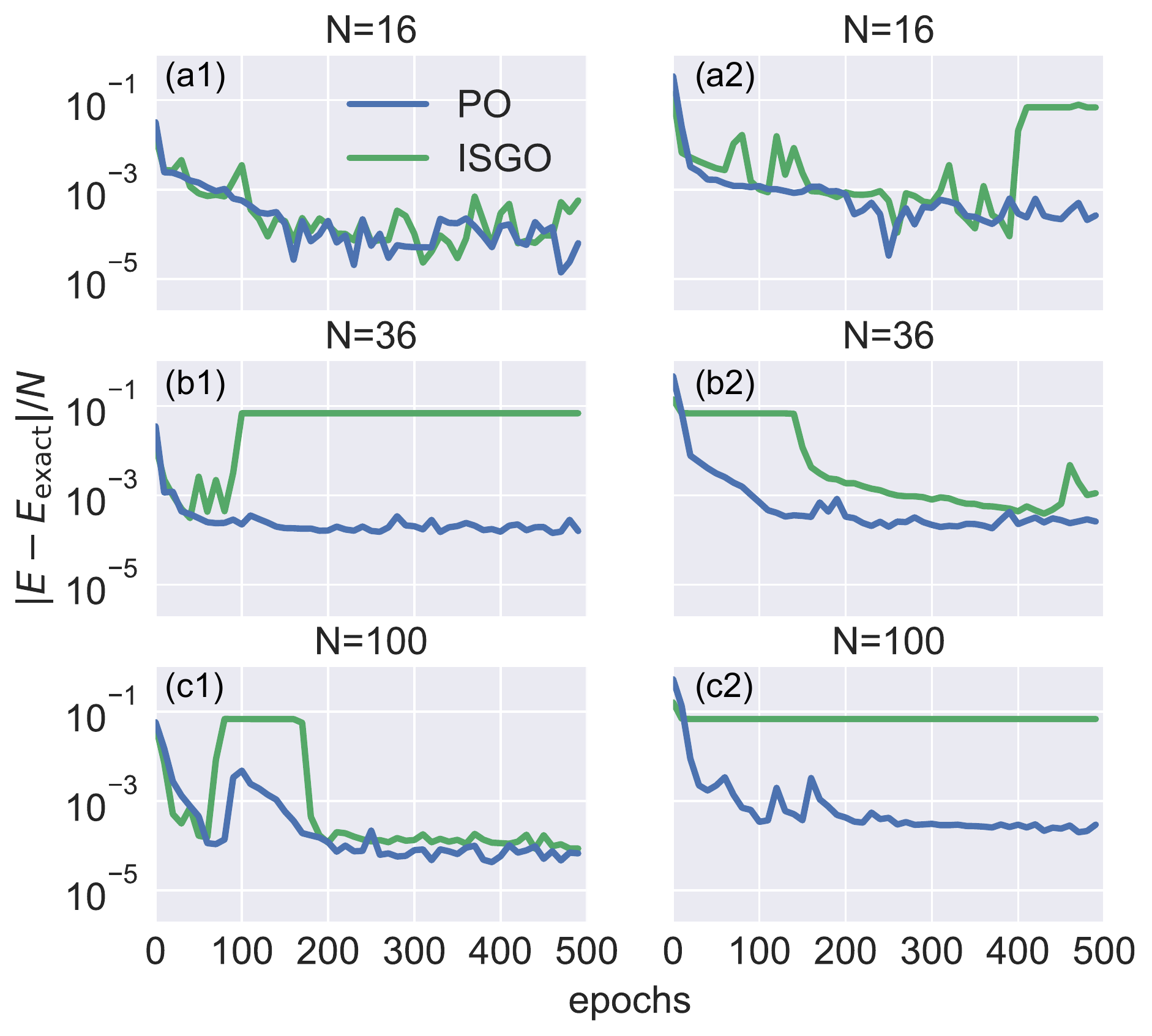}
    \caption{ Convergence comparison between the VMC-ISGO algorithm (green lines) and our VMC-PO algorithm (blue lines) for $h = -0.5$ (left panel) and $h = 0.5$ (right panel).
    Variational energies are compared to $E_{\rm exact}$ from exact diagonalization (ED) for $N=16$ or density matrix renormalization group (DMRG) for $N=36, 100$.
    All lines show 10-epoch moving averages.
    }
    \label{fig_tfim_05}
    \end{figure}

When $h < 0$, all elements of the ground state wave fucntion are positive real values
(up to a global phase) according to the Perron-Frobenius theorem~\cite{horn2013matrix}.
In this sense, the phase network that uniformly initialized is the exact solution,
since it outputs an identical phase for any sampled state $|\bm s^{(m)}\rangle$.
Therefore, the ground state of TFIM at $h=-0.5$ only depends on the amplitude network $\mathcal{A}$.
The left panel of Fig.~\ref{fig_tfim_05} shows the convergence of ground state energies during the training.
For a spin chain of $N=16$ [Fig.~\ref{fig_tfim_05}(a1)], both VMC-ISGO and VMC-PO can efficiently figure out the ground state.
For a longer spin chain, insufficient quantity of samples brings significant numerical instability
to the process of ground-state searching.
One of the worest consequences is being stucked at a local minimum,
and here the probable states of local minimum are the ground state of 
interating term in Eq.\,\eqref{ham_tfim}: $-\sum_i S^i_z S^{i+1}_z$,
i.e., the product state $\ket{\bm s}=\ket{\uparrow}^{\otimes N}$ or $\ket{\downarrow}^{\otimes N}$ 
with the degeneracy being twofold. 
The energy expectation in this case is $E = -0.25$ per site.
Only these two spin configurations are accessible as sampled states once the optimization falls into the local minimum.
It would result in a tremendous obstacle hindering the escape from the local minimum.
Without any constraints on parameters updating of the amplitude, 
VMC-ISGO may drive the NQS {\it Ans\"atze} into this local minimum as depicted by the flatten
plateau in green lines of Fig.~\ref{fig_tfim_05}(b1) and (c1).
In contrast, VMC-PO steadily drive the NQS {\it Ans\"atze} reaching the target ground state,
though numerical instabilities bring fluctuactions on the training process.

The robustness of VMC-PO is further demonstrated at $h = 0.5$.
When $h > 0$, the ground state wave function contains both positive and negative components whose signs are given by the phase network, which brings extra numerical instability to the optimization.
The training difficulty is thus dramatically enhanced, although the ground state at $h=0.5$ 
could simply be obtained via rotating all spins of the ground state at $h=-0.5$ by $\pi$ around the $S_z$ axis,
$|g(h=0.5)\rangle = \exp{-i\pi\sum_jS_z^j}|g(h=-0.5)\rangle$.
The progress of ground-state searching at $h=0.5$ is shown in the right panel of Fig.~\ref{fig_tfim_05}.
In this case, VMC-ISGO finally falls into the local minimum even in a $N=16$ spin chain.
When $N=100$, VMC-ISGO is stucked at this local minimum hopelessly.
On the other hand, VMC-PO keeps the same performance as the training of $h=-0.5$
whose relative error $|E - E_{\rm exact}|/E_{\rm exact} \lesssim 10^{-3}$.

\subsection{Heisenberg  Model}
Here, we consider the spin-$1/2$ Heisenberg antiferromagnetic model on 2D square lattice with wallpaper group $p4m$,
which consisting of translational, rotational and mirror symmetries.
The Hamiltonian of HAFM is,
\begin{equation}
    \label{ham_hsj1j2}
    H = J\sum_{\langle i,j \rangle} {\bf S}^i\cdot{\bf S}^j,
\end{equation}
where $J$ is the strength of nearest neighbour interaction.
Without loss of generality, we set $J = 1$ for antiferromagnetic nestest-neighbour interaction.
The amplitude and phase structures of ground states in 2D systems are much more complicated than the ones of 1D systems.
The VMC-ISGO approach is unstable in this case, 
hence we only implement our VMC-PO method to calculate the ground states in this section.

The phase structure of the ground state is described by Marshall sign rules (MSR)~\cite{marshall1955antiferromagnetism},
which gives a set of MSR basis $\{|\bm s^{(m)}\rangle'\}$,
\begin{equation}
    \label{eq:msr_basis}
    |\bm s^{(m)}\rangle' = \prod_{i\in A} 2S_z^i|\bm s^{(m)}\rangle = (-1)^{N_A^{\downarrow}}|\bm s^{(m)}\rangle,
\end{equation}
where $A$ is one of the two bipartite components of the square lattice
and $N_A^{\downarrow}$ is the number of spin-down particles in this sublattice.
The local spin operators in the MSR basis are,
\begin{equation}
    \label{eq:msr1}
    \left\{\begin{array}{llll}
    {S_x^i}' = - S_x^i, & {S_y^i}' = - S_y^i, & {S_z^i}' = S_z^i, & i\in A, \\
    {S_x^i}' = S_x^i, & {S_y^i}' = S_y^i, & {S_z^i}' = S_z^i, & i\in B,\end{array}\right.
\end{equation}
where $B$ is another checkerboard sublattice.
The $J$ term of Hamiltonian \eqref{ham_hsj1j2} in the MSR basis becomes,
\begin{equation}
    \label{eq:msr_ham}
    H' = \sum_{\langle i,j \rangle} S_z^i S_z^j - S_x^i S_x^j - S_y^i S_y^j,
\end{equation}
which makes all off-diagonal terms negative in the original basis $\{|\bm s^{(m)}\rangle\}$.
Perron-Frobenius theorem again ensures that all elements of the ground state of Hamiltonian \eqref{eq:msr_ham} are positive.
In other words, MSR gives the correct phase structures of the ground state.

\begin{figure} 
    \includegraphics[width=\columnwidth]{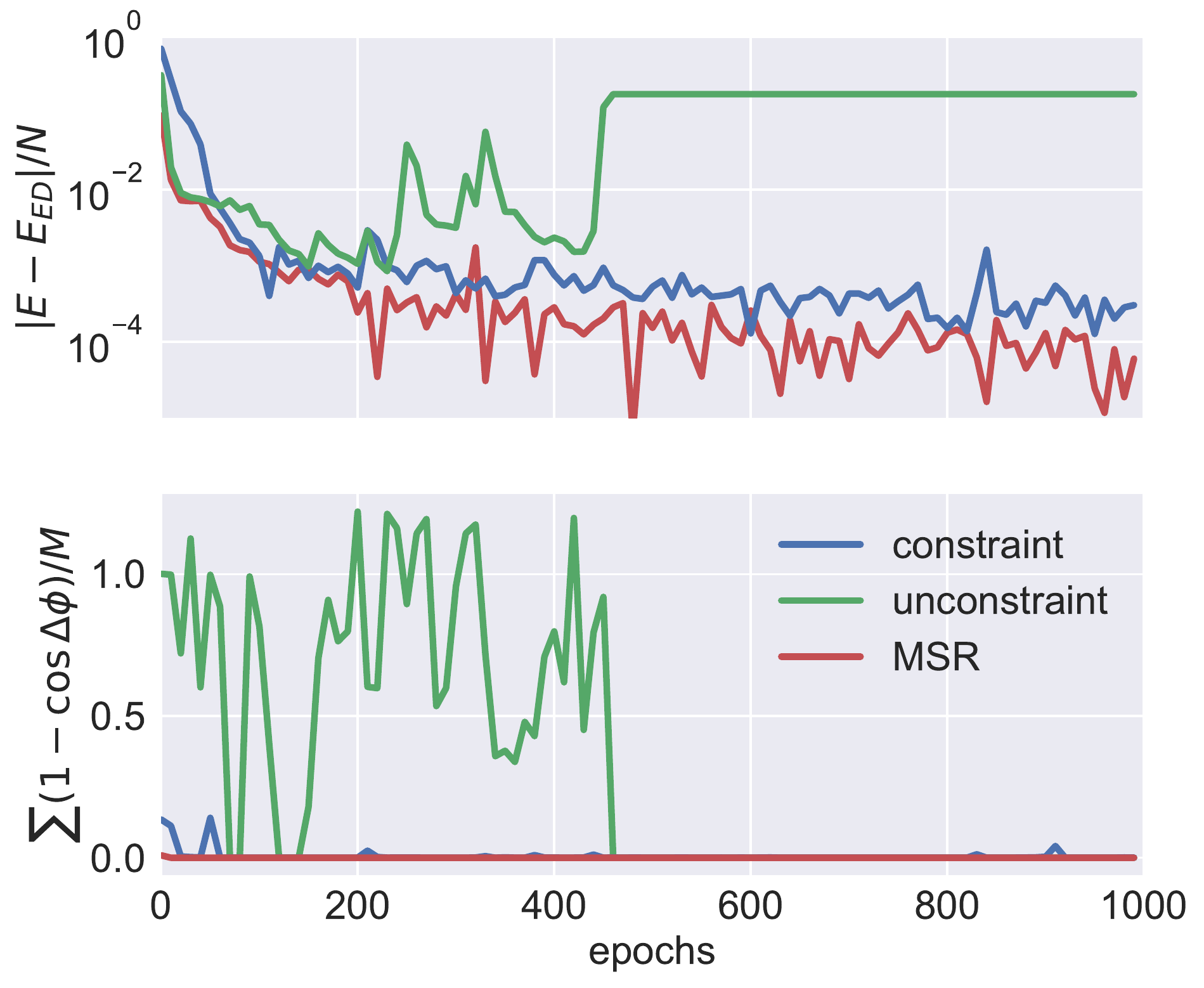}
    \caption{(a) Convergence comparison among phase constraint (blue line), phase unconstraint (green line),
    and using MSR (red line) for the 2D HAFM on a $6\times 6$ square-lattice.
    The MSR result is obatined by directly calculating the ground state energy of 
    Hamiltonian \eqref{eq:msr_ham} instead of embeding the MSR structures into neural networks.
    $E_{\rm ED}$ is the exact ground state energy from diagonalizing the Hamiltonian \eqref{ham_hsj1j2}~\cite{schulz1996}.
    (b) Mean phase variation of the sampled states.
    All lines show 10-epoch moving averages.
    }
    \label{fig_j20}
\end{figure}

Figure~\ref{fig_j20} shows the training process of a HAFM on a $6\times6$ square lattice.
Our VMC-PO results (blue line) from directly minimizing the objective $\mathcal{L}^{\rm clip}$ in Eq.\,\eqref{eq:ppo},
with both amplitude and phase networks, is quite close to the one using MSR (red line).
The phase structures from MSR is correct as the variants of phase $\Delta \Phi$ remaining
zero during training [as shown in Fig.\,\ref{fig_j20}(b)].
Without any constraints on the variation of phase updates [green line in Fig.~\ref{fig_j20}(a)],
the NQS finally falls into a local minimum and is trapped 
due to the instabilities existed during the parameters updating of phase network [as shown in Fig.\,\ref{fig_j20}(b)].
This local minimum is the ground state of $\sum_{\langle i,j\rangle}S_z^iS_z^j$,
the double degenerate N\'{e}el states (with energy $E=-0.5$ per site),
which is hard to escape due to the minor number of accessible samples.

\section{discusion and outlook} \label{sec_con}
In conclusion, we present a VMC-PO algorithm inspired by the 
reinforcement learning methods~\cite{ppo2017}. 
By utilizing the deep architectures of convolutional neural netwoks, 
the amplitude and phase of the complex-valued NQS {\it Ans\"atz} are composed and trained 
as two decoupled neural networks, respectively. 
The clipped Born probability and estimation of local energies in our VMC-PO algorithm are more constrained.
With samples reuse and only first-order gradient updates implemented, 
VMC-PO shows improved stability than poineering studies~\cite{yang2020}
during the training of a large number of parameters,
hence ensuring a deep neural-network representation.
We have benchmarked our algorithm with a 1D TFIM and 2D HAFM model on the square lattice,
the ground-state energies we obtained via VMC are competing with state-of-the-art results. 
With more powerful GPU resources to speedup the optimization, 
the VMC-PO would be promising to yield more precise ground-state representation.

Although the NQS representation with our VMC-PO algorithm shows competing ground-state energies searching
compared to the state-of-the-art computational methods,
such as the tensor network approaches concerning entanglement entropy, 
the physical insight here for NQS is a little bemused. 
We hope more physical insights can be interpreted from the future optimization for frustared quantum many-body state {\it Ans\"atze}.

\begin{acknowledgments}
We thank Dr.\;X.Y. Huang and Dr.\;J.J. Chen for offering the use of a NVIDIA RTX3090 GPU and a NVIDIA A100 GPU, respectively.
This work is supported by Open Research Fund Program of the State Key Laboratory of Low-Dimensional Quantum Physics (No.\;KF202111).
\end{acknowledgments}

\appendix

\section{training parameters} \label{app_A}
The parameters of network structure and the hyperparameters we use are shown
in Tab.\,\ref{table_Nparams} and \ref{table_Hparams}, respectively.
\begin{table}[htp!]
	\centering
	\caption{Network parameters}
	\label{table_Nparams}
	\begin{tabular}{c||c|c|c|c}
		\hline 
		& & & \\[-6pt]  
		Systems& \tabincell{c}{ Number of\\layers, $L$ } & \tabincell{c}{ Number of\\features, $F$ } & \tabincell{c}{ Size of\\kernel, $K$ } & $N_{\rm params}$ \\ 
		\hline
		& & & \\[-6pt]  
		TFIM ($N=16$)& 3& 8& 5 & 1504 \\
		\hline
        & & & \\[-6pt]  
		TFIM ($N=36$)& 3& 12& 7 & 4464 \\
		\hline
        & & & \\[-6pt]  
		TFIM ($N=100$)& 3& 16& 7 & 7744 \\
		\hline
        & & & \\[-6pt] 
		HAFM & 4& 4& 5$\times$5 & 19632\\
		\hline
	\end{tabular}
\end{table}

\begin{table}
	\centering
	\caption{Hyperparameters}
	\label{table_Hparams}
	\begin{tabular}{c||c}  
		\hline 
		&  \\[-6pt]  
		Name& Value \\ 
		\hline
		&  \\[-6pt]  
		clip ratio $\epsilon$& 0.1 \\
		\hline
        &  \\[-6pt]  
		$\beta_{\rm tg_{\max}}$& 0.2 \\
        \hline
        &  \\[-6pt]  
		$\beta_{\max}$& 10 \\
        \hline
        &  \\[-6pt]  
		$\beta_{\rm tg_{\min}}$& 0.05 \\
		\hline
        &  \\[-6pt]  
		$\beta_{\min}$& 0.5 \\
		\hline
        &  \\[-6pt]  
        learning rate & 2E-4 \\
		\hline
        &  \\[-6pt]  
        $n_{\rm opt}$& 50 \\
		\hline
	\end{tabular}
\end{table}

\bibliography{pporef}

\end{document}